**JOSÉ DÍAZ BEJARANO (1933-2019). UNA BIBLIOGRAFÍA**

*JOSÉ DÍAZ BEJARANO (1933-2019). A BIBLIOGRAPHY*

*José Manuel Vaquero*

Departamento de Física, Universidad de Extremadura

Resumen. José Díaz Bejarano (1933-2019) nació en Puebla de la Calzada (Badajoz). Estudió Química en la Universidad de Sevilla y Física en la Universidad de Madrid. Más tarde, desarrolló una exitosa carrera de investigación en varios centros internacionales incluyendo la Universidad de Hamburgo (Alemania), el CERN (Ginebra, Suiza) o la universidad Case Western Reserve (Cleveland, EEUU). Regresó a España a la Universidad de Extremadura, donde realizó una fecunda labor docente e investigadora en el Departamento de Física en la Facultad de Ciencias de Badajoz. En esta nota, se ofrece una bibliografía que contiene sus publicaciones, así como algunas noticias de su vida.

Palabras clave: Bibliografía, Historia de la Ciencia, Historia de la Física

Abstract. José Díaz Bejarano (1933-2019) was born in Puebla de la Calzada (Badajoz). He studied Chemistry in the University of Seville and Physics in the University of Madrid. Later, he developed a successful research career in several international centers including the University of Hamburg (Germany), CERN (Geneva, Switzerland) or Case Western Reserve University (Cleveland, USA). He returned to Spain to the University of Extremadura, where he did a successful teaching and research work in the Department of Physics (Faculty of Sciences at Badajoz). In this note, a bibliography is provided that contains his publications as well as some news from his life.

## 1. Introducción

El pasado 25 de febrero de 2019, moría en Badajoz D. José Díaz Bejarano. Personaje carismático entre los físicos de la Universidad de Extremadura (UEx), destacaba por su experiencia investigadora en centros extranjeros, así como por su generosidad y su fina ironía. Amante de la música, siempre llegaba con pequeños regalos y *souvenirs* para todo el personal del departamento de física tras sus exóticas vacaciones. Fui alumno suyo en el primer curso de la licenciatura en física (en la asignatura "Física General") y en el quinto (en la asignatura "Física Atómica"). Entre los alumnos de mi generación, la imagen de la mano de D. José con los dedos corazón, índice y pulgar formando un sistema de ejes cartesianos se popularizó como un icono de la física. Cuando en clase contaba alguna anécdota de su trabajo en Hamburgo o Cleveland, imaginábamos una vida idealizada llena de aventuras científicas que él mismo se encargaba de rebajar.

Si hay algo que debería ser destacado en especial es el hecho de que él estaba trabajando y publicando en centros y revistas de primer nivel de la física internacional en una época en la que Extremadura aún no contaba ni siquiera con una Facultad de Ciencias. Por ello, he creído de interés publicar esta bibliografía que recoge todos sus artículos en revistas científicas especializadas. También he incluido algunas notas esenciales sobre su biografía.

## 2. Algunas notas biográficas esenciales

D. José Díaz Bejarano nació en Puebla de la Calzada (Badajoz) el 14 de marzo de 1933 aunque pasó gran parte de su infancia y juventud en la ciudad de Badajoz. Realizó estudios universitarios en Sevilla, donde se licenció en Químicas en junio de 1955, y en Madrid, donde se licenció en Física en junio de 1958. Allí también se doctoró en Física en diciembre de 1960 con una tesis doctoral sobre un aspecto de la teoría de la relatividad dirigida por Julio Palacios, la última figura de la conocida como Edad de Plata de la ciencia española. En la Universidad Complutense de Madrid, en su Facultad de Ciencias, fue profesor ayudante de clases prácticas de "Física General" en el curso 1958-1959 y de "Análisis Dimensional" y de "Relatividad" en los cursos 1959-1960 y 1961-1962.

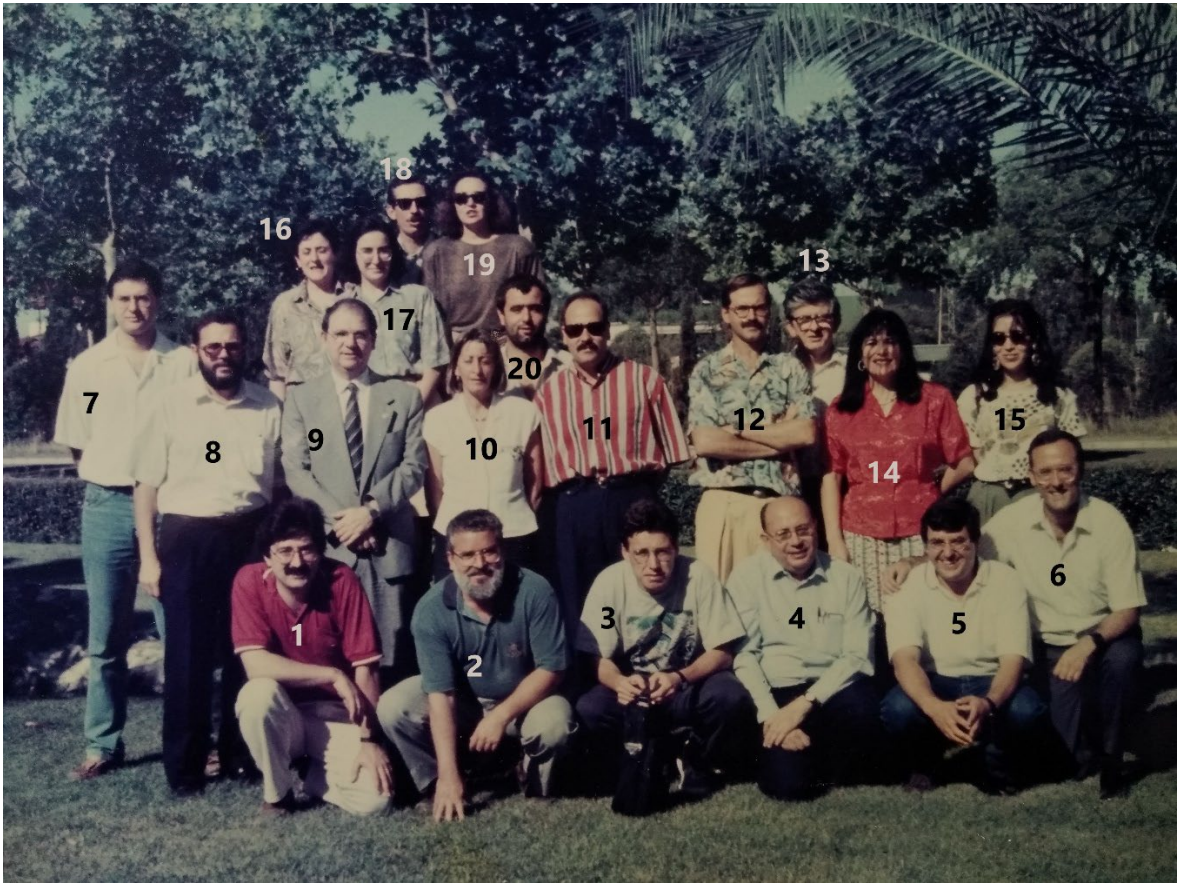

Figura 1. Miembros del Departamento de Física en el año 1993 fotografiados junto al conocido "edificio metálico" del campus universitario de Badajoz. Los nombres de los fotografiados son: (1) Juan José Morales Alcalá, (2) Francisco Cuadros Blázquez, (3) Vicente Garzó Puertos, (4) José Díaz Bejarano, (5) Juan Garrido Acero, (6) Francisco Luis Cumbrera Hernández, (7) Santos Bravo Yuste, (8) José Agustín García García, (9) José Morales Bruque, (10) Cristina Dorado Calasanz, (11) Feliciano Vera Tomé, (12) Andrés Santos Reyes, (13) Bronisław Jańczuk, (14) María Isabel Suero López, (15) María José Martín Delgado, (16) María Luisa González Martín, (17) María Luisa Cancillo Fernández, (18) José Margallo Guillén (19) María José Nuevo Sánchez y (20) Vidal Luis Mateos Masa.

A partir de este momento, se inicia una etapa de trabajo en el extranjero. En el curso 1960-1961, se incorpora como investigador asociado en el *II Physikalisches Institut und Desy*, asociado a la Universidad de Hamburgo (Alemania). Allí trabajaría hasta 1967, iniciando su investigación en física subatómica que cosecharía numerosos éxitos, publicando sus resultados en las más importantes revistas especializadas de la época. De 1968 a 1970, fue investigador asociado del CERN (*Conseil Européen pour la Recherche Nucléaire*) en

Ginebra (Suiza) pasando a la *Case Western Reserve University* en Cleveland (Ohio, USA) desde 1970 a 1972. Volvió a España, a la Junta de Energía Nuclear, durante el periodo 1973-1976. Su última estancia en el extranjero fue de nuevo en Ginebra de 1977 a 1978. Para una visión general de la ciencia en España durante esta época, pueden consultarse los últimos capítulos de la monografía de Sánchez Ron (2020).

A partir de este momento, D. José Díaz Bejarano intenta integrarse en la universidad española. En los cursos 1977-1978 y 1978-1979, es profesor adjunto interino de "Física Nuclear" de la Facultad de Ciencias de la Universidad de Sevilla. El 29 de septiembre de 1979, se convierte en Profesor Titular de "Física General" de la Facultad de Ciencias de la UEx. El 8 de abril de 1986, tomó posesión como Catedrático y en esta institución pasará ya el resto de su vida. Entre los cargos académicos que ostentó en la UEx, podemos citar la dirección de los Departamentos de Termodinámica y de Física (aunque en breves periodos). La Figura 1 muestra la mayor parte de los miembros del Departamento de Física en el año 1993 fotografiados junto al conocido "edificio metálico" del campus universitario de Badajoz. Por último, al final de su carrera, fue nombrado Profesor Emérito de la UEx. Tras una larga enfermedad, falleció en Badajoz el 25 de febrero de 2019.

### 3. Bibliografía

La mayor parte de la obra escrita de D. José Díaz Bejarano está publicada en revistas científicas especializadas. Se ofrece a continuación la lista completa de sus contribuciones:

Un análisis de esta bibliografía nos proporciona una imagen de su vida científica. Los tres primeros artículos están relacionados con la etapa temprana del trabajo de D. José Díaz Bejarano. El primero de sus artículos está relacionado con su inconclusa tesis para convertirse en Doctor en Química, que abandonó para dedicarse a la física. El segundo y el tercero están relacionados con el análisis dimensional, uno de los temas favoritos de Julio Palacios, su director de tesis en la Universidad Complutense. Es destacable que no publicara nada directamente relacionado con su tesis doctoral.

Los artículos 4-33, publicados en el periodo 1963-1981, corresponden a su primera etapa de madurez científica. Están publicados en importantes revistas internacionales. La autoría de estos artículos es múltiple como es habitual en este tipo de investigaciones de física subatómica donde se requieren enormes equipos experimentales para la realización de los complejos experimentos. Dado que D. José Díaz Bejarano estuvo trabajando en importantes centros de investigación, los artículos muestran una gran internacionalización de los trabajos, publicados en colaboración con importantes grupos de investigación extranjeros.

Quizás merezca la pena destacar sus colaboraciones con el grupo de la *Case Western Reserve University* (CWRU) en Cleveland (Ohio, USA). William Fickinger, uno de los colaboradores más activos de D. José, ha escrito recientemente una historia de la investigación en física hecha en la CWRU desde 1830 hasta la actualidad y ahí recuerda el paso de D. José por los laboratorios de esta institución americana en una sección dedicada a las resonancias multipión del capítulo 16 de su monografía. Ese capítulo está dedicado a la investigación experimental en física de partículas: "*Graduate student David Matthews and research associates Frank DiBianca (Carnegie Mellon), John Malko (Ohio University), and José Diaz Bejarano (CERN) were essential to the success of the program*" [El estudiante de posgrado David Matthews y los investigadores asociados Frank DiBianca (Carnegie Mellon), John Malko (Universidad de Ohio) y José Diaz Bejarano (CERN) fueron esenciales para el éxito del programa] (Fickinger, 2006, p. 259).

Los artículos 34-66, publicados desde 1982 hasta 2002, son el resultado de un enorme esfuerzo investigador realizado en la UEx. Cuando D. José Díaz Bejarano llega a esta universidad, se encuentra con una situación poco esperanzadora para un físico que ha trabajado con grandes equipos. No hay instrumentos ni laboratorios, pero sus amplios conocimientos matemáticos le proporcionan una oportunidad. Mientras los laboratorios se van conformando y los equipos van comprándose poco a poco, él y un grupo de jóvenes profesores (realizando sus tesis doctorales bajo su dirección) se dedicaron a utilizar las funciones elípticas para resolver numerosos problemas de la Física Matemática. El éxito es notable. En relativamente pocos años, se leen cuatro tesis doctorales (Tabla 1). Además, el ritmo de publicación en revistas especializadas internacionales se mantiene. Y eso fue gracias a estos trabajos de tipo teórico, mientras la investigación experimental mantuvo un constante crecimiento.

Tabla 1. Tesis doctorales dirigidas en la UEx por José Díaz Bejarano. Se lista el nombre del estudiante de doctorado, el título de la tesis y la fecha de lectura.

| Nombre del doctorado | Título de la tesis | Fecha de lectura |
|---|---|---|
| Alejandro Martín Sánchez | Aplicaciones de las funciones elípticas a los osciladores alineales | 23 junio 1983 |
| Conrado Miró Rodríguez | Estudio de sistemas con potenciales alineales mediante funciones elípticas | 22 junio1985 |
| Santos Bravo Yuste | Construcción de métodos de resolución aproximada de osciladores alineales usando funciones elípticas | 9 febrero 1990 |
| Ricardo Chacón García | Caos en osciladores y ecuaciones de onda no lineales bajo perturbaciones periódicas generalizadas | 19 septiembre 1995 |

Además de estos artículos en revistas, también se puede mencionar entre las publicaciones de D. José Díaz Bejarano su tesis doctoral, inédita. Esta llevaba por título "El experimento de Trouton y Noble según las nuevas fórmulas relativistas". Está fechada en junio de 1959 y fue defendida en la Universidad de Madrid (actual Universidad Complutense) bajo la dirección del profesor Julio Palacios, probablemente el físico español más importante de los años centrales del siglo XX (Oliva, 2013). En ella, se analiza el histórico experimento de Trouton y Noble (Butler, 1968; Teukolsky, 1996) a la luz de la nueva formulación de la teoría relativista que había realizado Julio Palacios. En su última etapa de trabajo, Palacios había hecho una profunda revisión de la teoría de la relatividad que no tuvo éxito (Sellés García, 1984). Fue en el marco de esta revisión donde José Díaz Bejarano realizó su tesis doctoral. Hasta donde sabemos, no publicó ninguno de los contenidos de su tesis, quizás porque las ideas de Julio Palacios sobre la relatividad fueron bastante heterodoxas. En cambio, sí publicaría trabajos (2-3) relacionados con otra de las líneas de investigación de Julio Palacios, el análisis dimensional. En este caso, el trabajo de Palacios fue reconocido internacionalmente y su libro sobre este tema (Palacios, 1956) fue traducido a diversos idiomas.

## 4. Comentarios finales

José Díaz Bejarano fue una persona singular en la formación de la primera comunidad de investigadores en física de la UEx. En estas páginas, se proporciona una bibliografía con sus artículos de investigación, así como una breve semblanza de su vida profesional. Un último hecho que puede ser de interés es que José Díaz Bejarano donó su biblioteca (con libros fundamentalmente de física y matemáticas) al Departamento de Física de la UEx donde actualmente se encuentra depositada.